\documentclass[12pt]{iopart}

\usepackage{iopams} 
\usepackage{graphicx}
\begin{document}

\title[]{Spin Hall conductance in a Y-shaped junction device in presence of tunable spin-orbit coupling}

\author{Sudin Ganguly$^1$\footnote{Corresponding 
author: sudin@iitg.ernet.in} and Saurabh Basu$^1$}

\address{$^1$Indian Institute of Technology Guwhati - Guwahati, Assam-781039, India}

\ead{saurabh@iitg.ernet.in}

\begin{abstract}
We study spin Hall effect in a three terminal Y-shaped device in presence
of tunable spin-orbit (SO) interactions via Landauer-B\"{u}ttiker formalism.  We have evolved a fabrication technique for creating different angular separation between the two arms of the Y-shaped device so as to investigate the effect of angular width on the spin Hall conductance (SHC). A smaller angular separation yields a larger conductance. Also arbitrary orientation of the spin quantization axes yields interesting three dimensional contour maps for the SHC corresponding to different angular separation of the Y-shaped device. The results explicitly show breaking of the spin rotational symmetry. Further a systematic study is carried out to compare and contrast between the different SO terms, such as Rashba and Dresselhaus SO interactions and the interplay of the angular separation therein.
\end{abstract}

\vspace{2pc}

\maketitle

\section{Introduction}
After prediction and detection of the spin Hall effect \cite{dyakonov1,dyakonov2,kato,sih} (SHE), the spin-dependent electronic transport has been a central focus of investigation in mesoscopic physics because of its possible applications to spintronics \cite{wolf,zutic}. Generation of dissipationless spin current \cite{murakami} is one of the features that is believed to be crucial in this respect. In early attempts, generation of spin-polarized currents were obtained by attaching ferromagnetic metallic contacts to the semiconductors \cite{wolf,dutta-das}. But, the efficiency of the spin injection from a ferromagnet into a semiconductor is poor because of the conductivity mismatch \cite{Schmidt} between the two. This drawback can be overcome by producing spin-polarized current intrinsically. Here comes the role of spin-orbit (SO) interaction. A strong spin-orbit scattering generates spin-polarized electrons intrinsically \cite{l-l}.

In general, two types of spin-orbit coupling terms can be present in semiconductor heterostructures having two dimensional electron systems (2DES). One of them is the Dresselhaus spin-orbit coupling (DSOC) which originates from the inversion asymmetry of the zinc blende type of structures \cite{dresselhaus} . The other is the Rashba spin-orbit coupling (RSOC) which originates due to the effective electric field originating from the asymmetry of the potential confining the 2DES \cite{rashba}. The Dresselhaus term is found to be dominant in large band gap materials and the strength can be controlled easily by tunning the quantum well width \cite{lommer}. On the other hand, the Rashba term is dominant in narrow-gap systems where the strength of the Rashba term can be controlled by external gate voltages \cite{nitta,engels}. The interplay of both types of spin-orbit coupling on the conductance characteristics of nanostructures has been investigated both theoretically \cite{mish,loss,shen,moca} and experimentally \cite{park,sasaki}.

Geometry of the scattering region also plays an important role in order to study effects of spin-polarization in presence of spin-orbit interaction. Four terminal junction devices have been studied where unpolarized charge current is driven through the  longitudinal leads attached to a semiconducting region with SO coupling induces a pure spin current at the transverse voltage  probes without accompanying any charge current \cite{nico1,ting}. In particular, three terminal structures such as T-shaped \cite{kramer}, Y-shaped \cite{pareek,cumm1,cumm2,wojcik} devices have also been studied in presence of spin-orbit interaction. In a three terminal structure, one terminal acts as an input to the device, through which unpolarized charge current is injected into the device. The other two terminals act as outputs through which the spin-polarized currents flow out of the device. 

Since a three terminal structure is a suitable candidate for studying the SHE, in this paper, we have studied the behaviour of a special type of three terminal device with Y-shaped structure.Since the angular separation between the arms of the Y-shaped geometry can be relevant for studying SHE, we have considered different angles as shown in Fig.\ref{setup}. Also because the rotational symmetry is broken in spin space in presence of the spin-orbit interaction, the spin quantization axes also play an important role in the context of SHE. Motivated by these, we have studied the effects of the angular variation and orientation of the spin quantization axes on the spin Hall conductance of such a Y-shaped junction with tunable SO interactions.

We organize our paper as follows. In the following section, we present a prescription of fabricating a Y-shaped device. The theoretical formalism leading to the expression for the spin Hall conductance using Landauer-B\"{u}ttiker formula are presented in the next section. After that we include an elaborate discussion of the results obtained for the spin Hall conductance in presence of the SO interaction. We have included an interesting comparison for the conductance properties in presence of Rashba vis-a-vis Dresselhaus SO interactions.

\section{Fabrication of Y-shaped devices}
\label{fyd}
We begin our discussion by a prescription of fabricating a Y-shaped
junction device which should be interesting from an experimental
perspective.

%\begin{wrapfigure}{r}{0.5\textwidth} 
\begin{figure}
\begin{center}
\fbox{\includegraphics[width=0.3\textwidth]{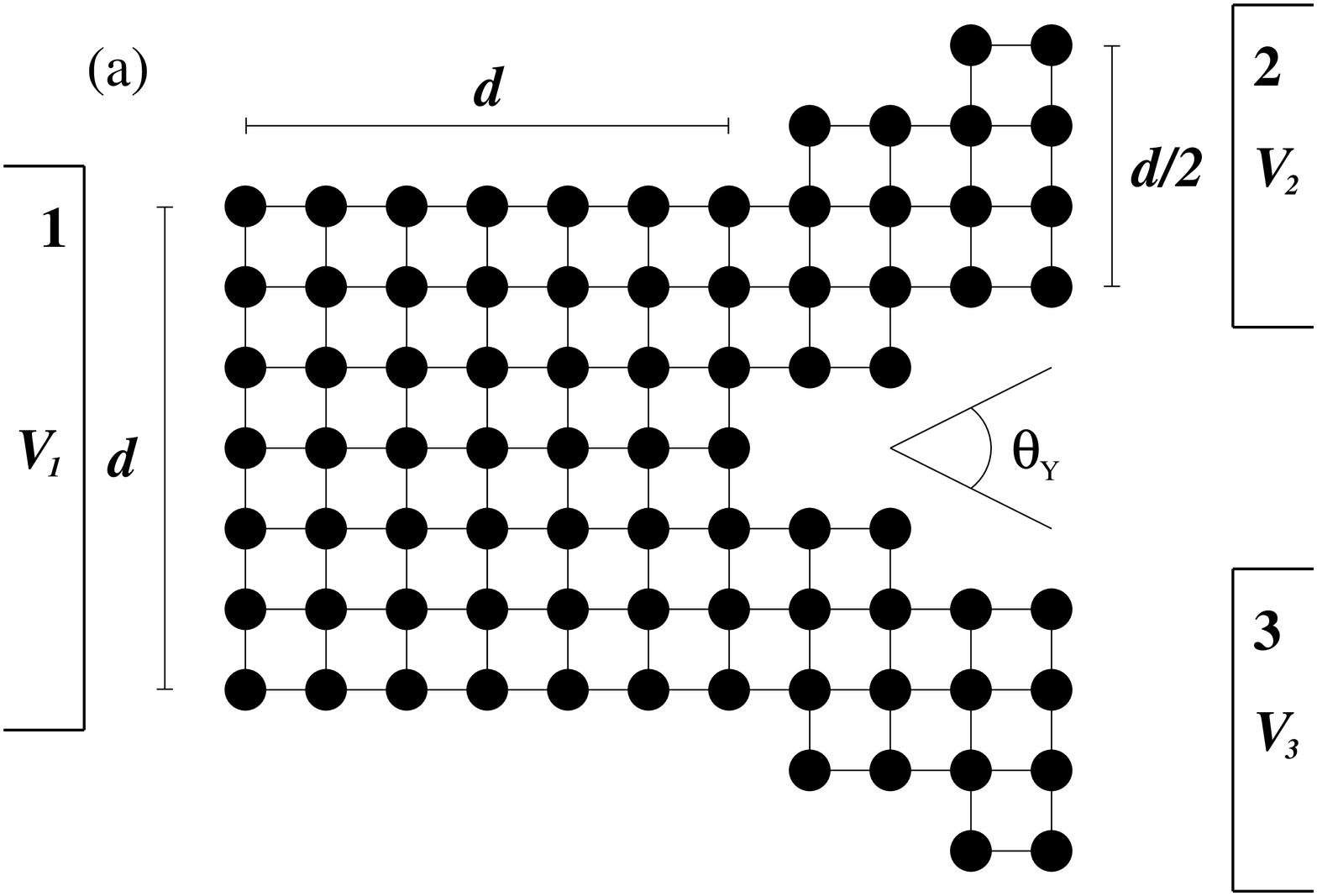}}
\fbox{\includegraphics[width=0.3\textwidth]{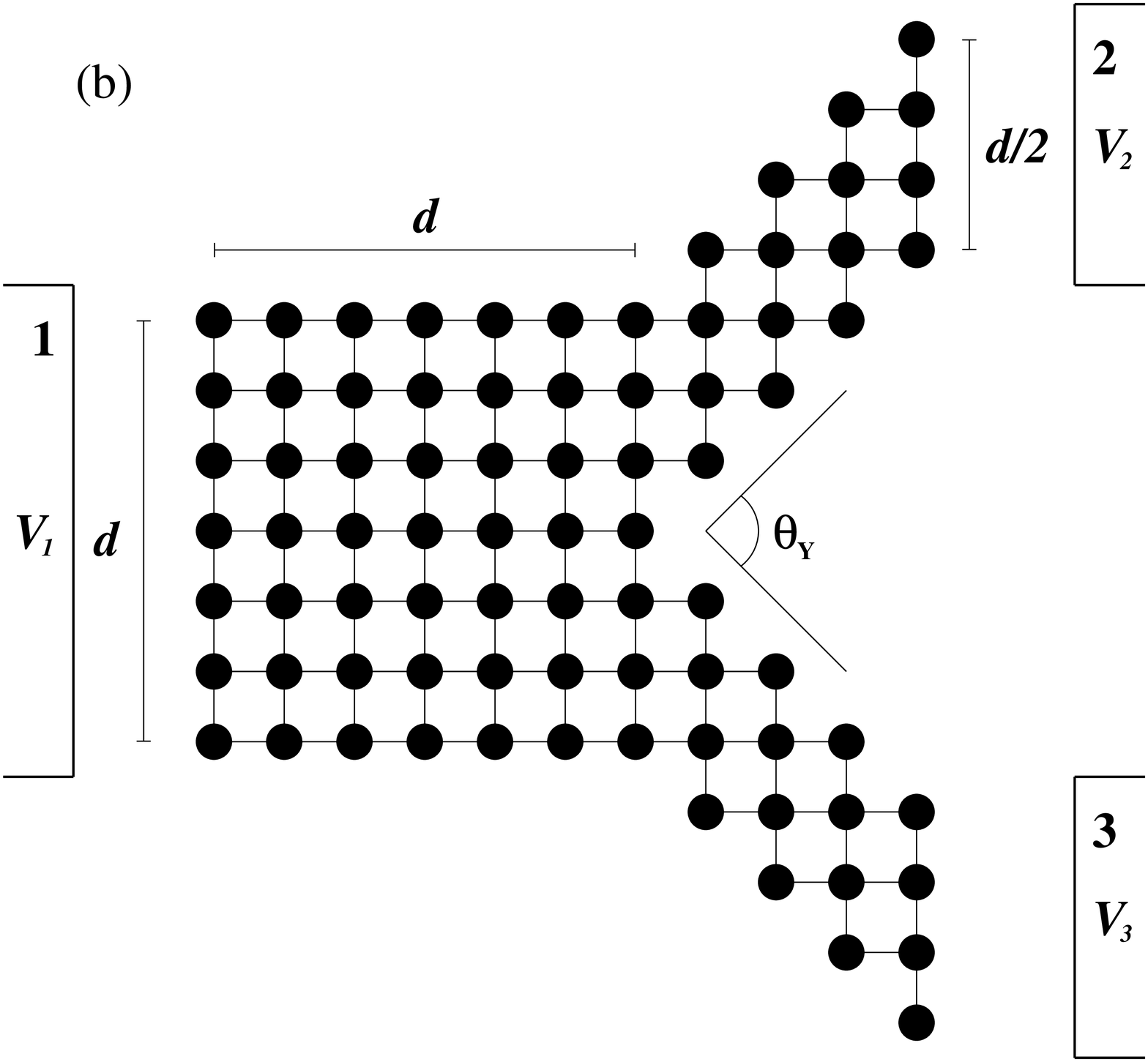}}
\fbox{\includegraphics[width=0.3\textwidth]{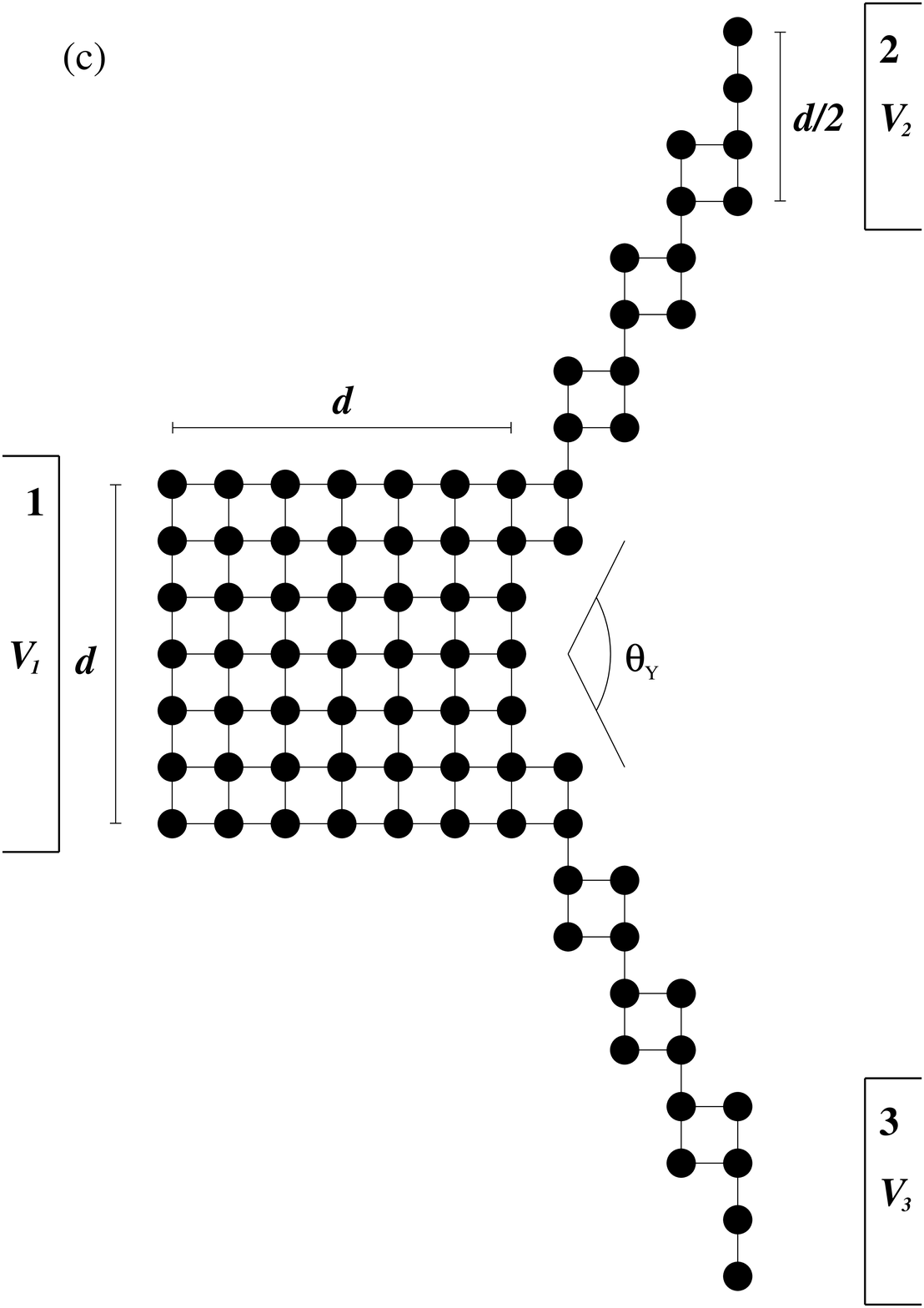}}
\caption{Y-shaped three terminal junction devices with three
  different angles $\left(\theta_Y\right)$. $V_1$, $V_2$ and $V_3$ are
  the applied voltages at the three terminals. The leads are not shown
  in the figure.}
\label{setup}
\end{center}
\end{figure}
%\end{wrapfigure}

We choose a three-probe measuring set-up as shown in Fig.\ref{setup}
to observe the spin Hall effect. Here the three ideal semi-infinite
leads are attached to the central conducting region, which in our case
is the Y-shaped device having a square lattice geometry and
includes spin orbit interaction. The leads denoted
by 1, 2 and 3 are semi-infinite in nature. The voltages applied at the
leads are $V_1$, $V_2$ and $V_3$ respectively. The width of the
scattering region is $d$, while the arms has width $d/2$. In this work, we have taken three
different Y-shapes by changing the angle between the two arms of the
Y, and call it $\theta_Y$ as shown in Fig.\ref{setup}. 
\begin{figure}
\fbox{\includegraphics[scale=0.4]{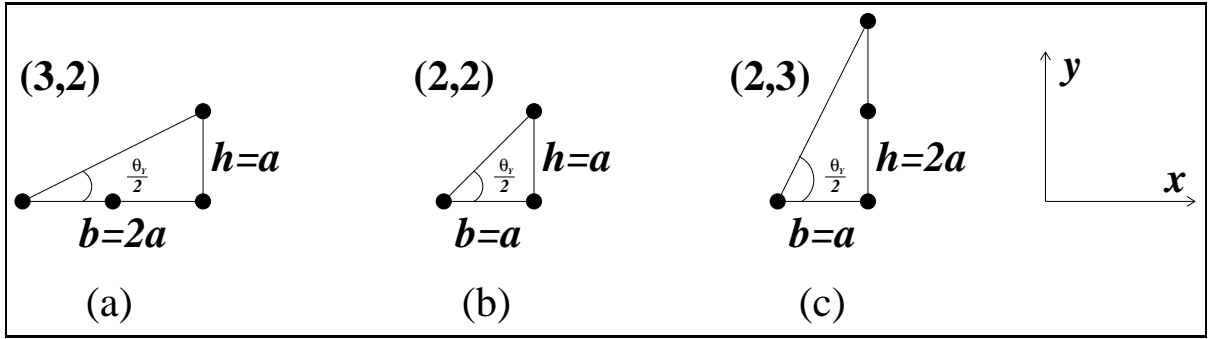}}
\caption{Measurement of the angle between the two arms of the Y-shaped
  device is shown. $b$ is the base of the triangle and $h$ is the
  height. $a$ is the lattice constant. According to Fig.\ref{setup},
  this angle is half of $\theta_Y$ as depicted in the given figure.}
\label{angle}
\end{figure}

Fig.\ref{angle} provides a technique of how one can fix the angle,
$\theta_Y$. According to Fig.\ref{setup}, $\theta_Y$ is twice the
angle as shown in Fig.\ref{angle}. For the Y-shape shown in
Fig.\ref{setup}(a), we add the lattice sites in the arms of Y as
depicted in Fig.\ref{angle}(a). First we add three sites along $x$-axis
with spacing `$a$' and then add another site along $y$-axis just above
the third site along $x$. We repeat the same procedure to build up the
rest of the arm of the Y-geometry. Since we need three sites along
$x$-axis and two sites along $y$-axis we call it a (3,2) scheme. The
calculation of the angle in now straight forward from the
geometry. From Fig.\ref{angle}, $b$ is the base of the triangle and
$h$ is the height. In the (3,2) scheme, $b=2a$ and $h=a$. Hence, the
angular separation between the arms of the Y will be twice the
calculated angle and is, $\theta_Y=2\tan^{-1}{\frac{a}{2a}}=53.13^\circ$. Similarly, corresponding to $\theta_Y=90^\circ$, we need the (2,2) scheme, for which two sites along $x$-axis and one site along $y$-axis
above the second site along $x$ are required, as shown in
Fig.\ref{angle}(b). In the given case,
$\theta_Y=2\tan^{-1}{\frac{a}{a}}=90^\circ$. To obtain an angle, $\theta_Y$ greater than $90^\circ$, we adopt (2,3) scheme as shown in Fig.\ref{angle}(c). Here, $\theta_Y=2\tan^{-1}{\frac{2a}{a}}=128.87^\circ$.

However, we have been able to generate a number
of other values for the angular separation, $\theta_Y$ by following
the prescription given above.
\section{Theoretical formulation}
\label{sys-ham}
\subsection{System and Hamiltonian}
The single particle Hamiltonian for an electron in presence of both Rashba and Dresselhaus spin-orbit interaction in a two dimensional electron system is given by,
\begin{equation}
H = \frac{\mathbf{p}^2}{2m^*} + \frac{\alpha}{\hbar}\left(\sigma_xp_y - \sigma_yp_x\right) + \frac{\beta}{\hbar}\left(\sigma_xp_x - \sigma_yp_y\right)
\label{continum}
\end{equation}
where $\mathbf{p}$ $(=p_x,\;p_y)$ is the two dimensional momentum operator, $m^*$ is the effective mass, $\sigma_x$ and $\sigma_y$ are the components of the Pauli matrices. $\alpha$ and $\beta$ respective denote the Rashba and Dresselhaus spin-orbit coupling strengths.

We discretize the Hamiltonian via a tight binding approximation with nearest neighbour hopping. The resulting Hamiltonian becomes,

\begin{eqnarray}
H = \epsilon\sum\limits_{i,\sigma} c_{i\sigma}^{\dagger}
c_{i\sigma} + t\sum\limits_{\langle ij\rangle,\sigma}
c_{i\sigma}^{\dagger} c_{j\sigma}  \nonumber\\+V_R\sum\limits_{i} 
\left[\left(c_{i\uparrow}^{\dagger} c_{i+\delta_x \downarrow} -
  c_{i\downarrow}^{\dagger} c_{i+\delta_x \uparrow}\right)   -i
  \left(c_{i\uparrow}^{\dagger} c_{i+\delta_y \downarrow} +
  c_{i\downarrow}^{\dagger} c_{i+\delta_y \uparrow}\right) \right]\nonumber\\
+V_D\sum\limits_{i} \left[(-i)\left(c_{i\uparrow}^{\dagger} c_{i+\delta_x \downarrow} +
  c_{i\downarrow}^{\dagger} c_{i+\delta_x \uparrow}\right)   +
  \left(c_{i\uparrow}^{\dagger} c_{i+\delta_y \downarrow} -
  c_{i\downarrow}^{\dagger} c_{i+\delta_y \uparrow}\right) \right]
\label{h}
\end{eqnarray}

Here $\epsilon$ is the on-site potential and $t=\hbar^2/2m^*a^2$ is the hopping
strength, $V_R=\alpha/a$ and $V_D=\beta/a$ are the Rashba and Dresselhaus coupling
strengths respectively, $a$ being the lattice constant. $\delta_{x/y}$ is the
unit vector along $x/y$ direction.

It is assumed that the Rashba and Dresselhaus SO interactions are present only in the Y-shaped device. The leads are metallic and semi-infinite in nature. The leads are free from any kind of SO interactions so as to avoid any kind or spin flips at the boundaries.

\subsection{Formulation spin Hall conductance}
Since the rotational symmetry in spin space is lost in presence of spin-orbit interaction, the quantization axes of the spin play a vital role in measuring spin current. Hence we choose the spin quantization axis along an arbitrary direction, say $\hat{\mathbf{u}}$, pointing along $(\theta$, $\phi)$, where $\theta$ and $\phi$ are the usual spherical angles $(\hat{\mathbf{u}}=\sin{\theta}\cos{\phi},\sin{\theta}\sin{\phi,\cos{\theta}})$. 

Now we proceed to evaluate the expression for spin Hall conductance. In order to get pure spin current, we treat terminal 2 as a voltage probe (Fig.\ref{setup}). As a result pure spin current will flow through terminal 2, due to the flow of charge current between terminals 1 and 3. 
For the three terminal case, the spin Hall
conductance is defined as \cite{mou},

\begin{equation}
G_{SH} = \frac{\hbar}{2e}\frac{I_2^s}{V_2-V_1}
\label{def-gls}
\end{equation}
where $I_2^s$ is the spin current flowing
through lead-2. $V_m$ is the potential at
the $m$-th lead.

The calculation of the electric and spin currents is based on the
Landauer-B\"{u}ttiker multi-probe formalism
\cite{Buttiker}. The charge and spin currents flowing through lead
$m$ $(m=1,2,3)$ with potential, $V_m$ can be written in terms of the spin
resolved transmission probability as \cite {pareek},

\begin{equation}
I_m^q= \frac{e^2}{h} \sum\limits_{n\neq m,\sigma,\sigma^\prime}
\left(T_{nm}^{\sigma\sigma^\prime}V_m
-T_{mn}^{\sigma^\prime\sigma}V_n\right)
\label{iq}
\end{equation}
and,
\begin{eqnarray}
I_m^s = \frac{e^2}{h} \sum\limits_{n\neq m,\sigma^\prime}
\left[\left(T_{nm}^{\sigma^\prime\sigma} -
  T_{nm}^{\sigma^\prime-\sigma}\right)V_m \nonumber \right.\\\left.+
  \left(T_{mn}^{-\sigma\sigma^\prime} -
  T_{mn}^{\sigma\sigma^\prime}\right)V_n\right] \nonumber \\=
\frac{e^2}{h}\sum\limits_{n\neq m} \left[T_{nm}^{out} V_m -
  T_{mn}^{in} V_n\right]
\label{is}
\end{eqnarray}

where, we have defined two useful quantities as follows,
\begin{eqnarray}
T_{pq}^{in} &=& 
T_{pq}^{\uparrow\uparrow} +
T_{pq}^{\uparrow\downarrow} - T_{pq}^{\downarrow\uparrow} -
T_{pq}^{\downarrow\downarrow} \nonumber\\ 
T_{pq}^{out} &=&
T_{pq}^{\uparrow\uparrow} + T_{pq}^{\downarrow\uparrow} -
T_{pq}^{\uparrow\downarrow} - T_{pq}^{\downarrow\downarrow}
\end{eqnarray}

Physically, the term $\frac{e^2}{h}\sum\limits_{n\neq m} T_{nm}^{out}
V_m$ is the total spin current flowing out from the $m$-th lead with
potential $V_m$ to all other $n$ leads, while the term
$\frac{e^2}{h}\sum\limits_{n\neq m} T_{mn}^{in} V_n$ is the total spin
current flowing into the $m$-th lead from all other $n$ leads having
potential $V_n$.

The zero temperature conductance, $G_{pq}^{\sigma\sigma^\prime}$ that
describes the spin resolved transport measurements, is related to the
spin resolved transmission coefficient by \cite
{land_cond,land_cond2},

\begin{equation}
G_{pq}^{\sigma\sigma^\prime} = \frac{e^2}{h} T_{pq}^{\sigma\sigma^\prime}(E)
\end{equation}

The transmission coefficient can be calculated from
\cite{caroli,Fisher-Lee},

\begin{equation}
T_{pq}^{\sigma\sigma^\prime} = {\rm{Tr}}\left[\Gamma_p^\sigma G_R
  \Gamma_q^{\sigma^\prime} G_A\right]
\end{equation}
$\Gamma_p^\sigma$'s are the coupling matrices representing the
coupling between the central region and the leads, and they are
defined by the relation \cite{dutta},
\begin{equation}
\Gamma_{p}^{\sigma} = i\left[\Sigma_p^\sigma -
  (\Sigma_p^\sigma)^\dagger\right]
\end{equation} 
Here $\Sigma_p^\sigma$ is the retarded self-energy for spin $\sigma$ associated with the
lead $p$. The self-energy contribution is computed by modeling each
terminal as a semi-infinite perfect wire \cite{nico}.

The retarded Green's function, $G_R$ is computed using
\begin{equation}
G_R = \left(E - H - \sum\limits_{p=1}^4 \Sigma_p\right)^{-1}
\end{equation} 
where $E$ is the Fermi energy and $H$ is the model
Hamiltonian for the central conducting region as given in Eq.(\ref{h}).  $G_A$ is the advanced
Green's function and is given by,
\begin{equation}G_A= G_R^\dagger
\end{equation}

Now, following the spin Hall phenomenology, in our set-up since lead-2
is a voltage probe, $I_2^q = 0$. Also, as the currents in various
leads depend only on voltage differences among them, we can set one of
the voltages to zero without any loss of generality. Here we set $V_1
= 0$ and $V_3=1$. With the help of these conditions, from
Eq.(\ref{iq}), one can determine the voltage, $V_2$,
\begin{equation}
V_2 = \frac{T_{23}}{T_{12}+T_{32}}
\end{equation}
Further the spin current flowing through terminal $2$ (from Eq.(\ref{is})) is,
\begin{eqnarray}
I^s_2 =
\frac{e^2}{h}\left[\left(T^{\mathrm{out}}_{12}+T^{\mathrm{out}}_{32}\right)V_2
  - T^{\mathrm{in}}_{23}\right]
\end{eqnarray}
Finally, from Eq.(\ref{def-gls}) the expression for the spin Hall conductance is given by,
\begin{eqnarray}
 G_{SH} =
 \frac{e}{4\pi} \left[ \left(T_{12}^{\mathrm{out}} +
   T_{32}^{\mathrm{out}}\right) -
   T_{23}^{\mathrm{in}}\frac{T_{12}+T_{32}}{T_{23}}\right]
\end{eqnarray}

\section{Results and discussion}
We have investigated the effects of the angle variation of the
Y-shaped junction in presence of Rashba  and Dresselhaus SO couplings on the
experimentally measurable quantity, namely the spin Hall conductance ($G_{SH}$). We have also studied the effect of the orientation of the quantization axis of spin on the spin Hall conductance.

We briefly describe the values of different parameters used in our
calculation. Throughout our work, we have considered for the Y-shaped
system, $d=20a$ (see Fig.\ref{setup}), onsite term, $\epsilon = 0$, hopping term, $t=1$. All the energies are
measured in unit of $t$. Further we choose a unit where $c=h=e=1$. The spin Hall conductance, $G_{SH}$ is measured in
units of $\frac{e}{4\pi}$. Also the lattice constant, $a$ is taken to be
unity.  For most of our numerical calculations we have used KWANT
\cite{kwant}. 

From the experimental perspective, we have included a brief discussion on the realistic values of the SO couplings, observed in materials. In GaAs, the effective mass, $m^*=0.067m_0$ and the lattice constant, $a=0.5653$ nm. With these values, the hopping integral becomes $t\simeq$ 1.8 eV (from the discussion following Eq.(\ref{h})). Also in InAlAs/InGaAs it is found that the Rashba parameter is, $\alpha\sim0.67  \times 10^{-11}$ eV-m \cite{nitta,park}. Then in our case, $V_R=\alpha/a\simeq0.01$ eV. Since we are denoting all the energy units in terms of $t$, $V_R/t\simeq0.006$. Which is pretty small. However, recently, in topological insulators such as Bi$_2$Se$_3$, the Rashba coupling parameter is found out to be $\sim4 \times 10^{-10}$  eV-m \cite{king}, polar semiconductor such as BiTeI shows a bulk Rashba coupling parameter $\sim3.85 \times 10 ^{-10}$  eV-m \cite{ishi}. With these higher values of $\alpha$, we have $V_R/t\simeq0.4$, which is of the order of unity and precisely similar in magnitude to what have been used in our work. In fact, we have considered the Rashba and Dresselhaus coupling parameters in the interval $[0:1]$.

In this work, we have taken three different angles for
the Y-shaped device, such as, $\theta_Y$ values to be
less than, equal to and greater than $90^\circ$. in particular we have considered, $\theta_Y=53.13^\circ,\; 90^\circ\;
\mathrm{and} \; 128.87^\circ$ as elaborated earlier.

\begin{figure*}
\begin{center}
\includegraphics[width=0.33\textwidth]{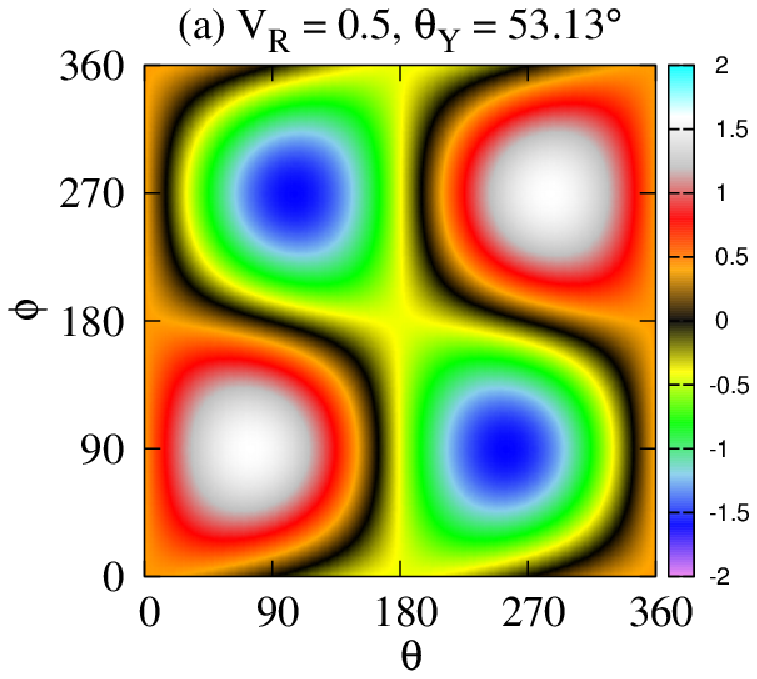}\includegraphics[width=0.33\textwidth]{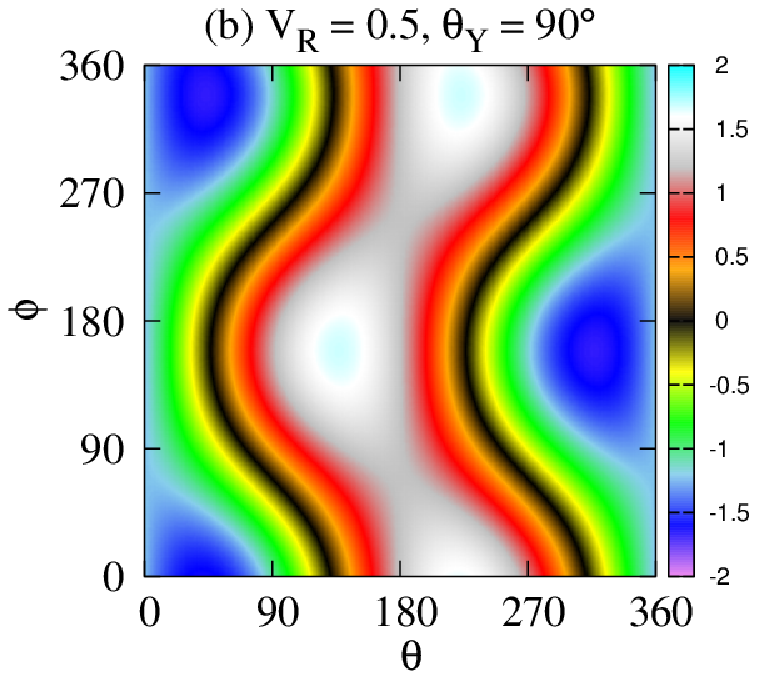}\includegraphics[width=0.33\textwidth]{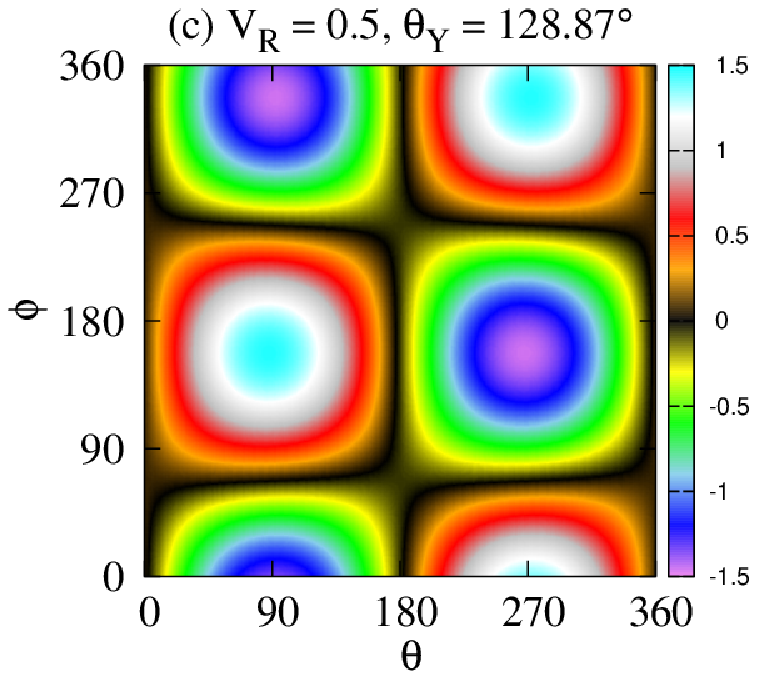}
\caption{(Color online) Spin Hall conductance, $G_{SH}$ is plotted as a function of the parameters describing the spin quantization axes, namely $\theta$ and $\phi$ for a Fermi energy $E=-2t$ in presence of Rashba spin-orbit coupling with strength, $V_R=0.5$.}
\label{gs_theta_phi}
\end{center}
\end{figure*}

We study the behaviour of the spin Hall conductance as a function of the spin quantization axes parameter, $\theta$ and $\phi$. In Fig.\ref{gs_theta_phi}, we show the variation of $G_{SH}$ as a function of the spin quantization axes in presence of Rashba SO coupling with strength, $V_R=0.5$ for three different angles of the Y-shaped device. We set the Fermi energy, to be at $E=-2t$. The nature of $G_{SH}$ for the three plots in Fig.\ref{gs_theta_phi} are clearly distinct from one another though the RSOC strength is the same. This is because of the difference in the angle, $\theta_Y$, which introduces different scattering environment for the electrons. 

In Fig.\ref{gs_theta_phi}(a), $G_{SH}$ shows a symmetric nature along the $\theta=\phi$ line for $\theta_Y=53.13^\circ$. The color map in Fig.\ref{gs_theta_phi}(a) can be divided into four regions, namely I. ($\theta: 0^\circ-180^\circ$, $\phi: 0^\circ-180^\circ$), II. ($\theta: 0^\circ-180^\circ$, $\phi: 180^\circ-360^\circ$), III. ($\theta: 180^\circ-360^\circ$, $\phi: 0^\circ-180^\circ$) and IV. ($\theta: 180^\circ-360^\circ$, $\phi: 180^\circ-360^\circ$). At the center of the each region, $G_{SH}$ has a periodic behaviour (circular patches) which can be seen from the coloured circles. This is raminiscent of the phase space plot for a simple harmonic oscillator. In Fig.\ref{gs_theta_phi}(b), $G_{SH}$ shows different behaviour as a function of $\theta$ and $\phi$ for $\theta_Y=90^\circ$. For fixed values of $\phi$, for lower values of $\theta$, $G_{SH}$ starts form negative values. It gradually increases to zero as $\theta$ increases. Finally in the vicinity of $\theta=180^\circ$, $G_{SH}$ becomes positive emphasizing the spin rotational broken symmetry state. For $\theta_Y=128.87^\circ$, the behaviour of $G_{SH}$ is completely different form the previous two plots as shown in Fig.\ref{gs_theta_phi}(c). Here we get few bounded regions and each region is separated by zero $G_{SH}$ as shown by the dark black line. Along the three lines, namely $\theta=0^\circ$, $\theta=180^\circ$ and $\theta=360^\circ$, the value of $G_{SH}$ is zero. An additional observation is that the magnitude of the spin Hall conductance is lower than the previous two cases.

\begin{figure*}[h]
\begin{center}
\includegraphics[width=0.33\textwidth]{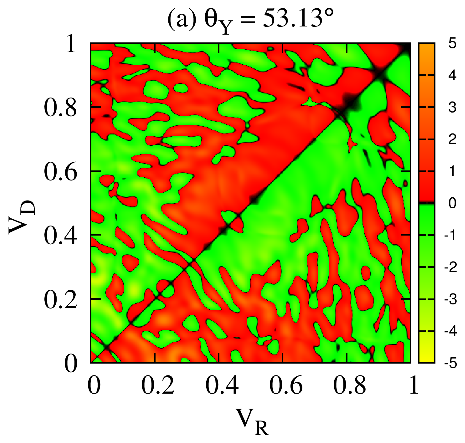}\includegraphics[width=0.33\textwidth]{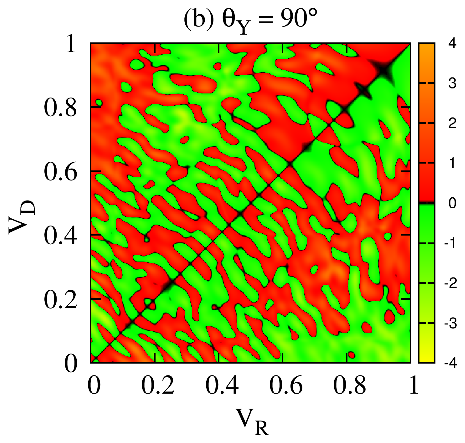}\includegraphics[width=0.33\textwidth]{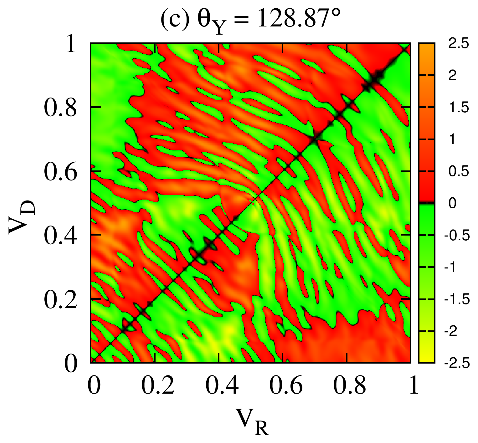}
\caption{(Color online) Spin Hall conductance, $G_{SH}$ is plotted as a function of the spin-orbit interaction strengths, $V_R$ and $V_D$ for a Fermi energy $E=-2t$. A distinct (and familiar) antisymmetric behaviour is noted.}
\label{gs_vr-vd}
\end{center}
\end{figure*}
\begin{figure}[h]
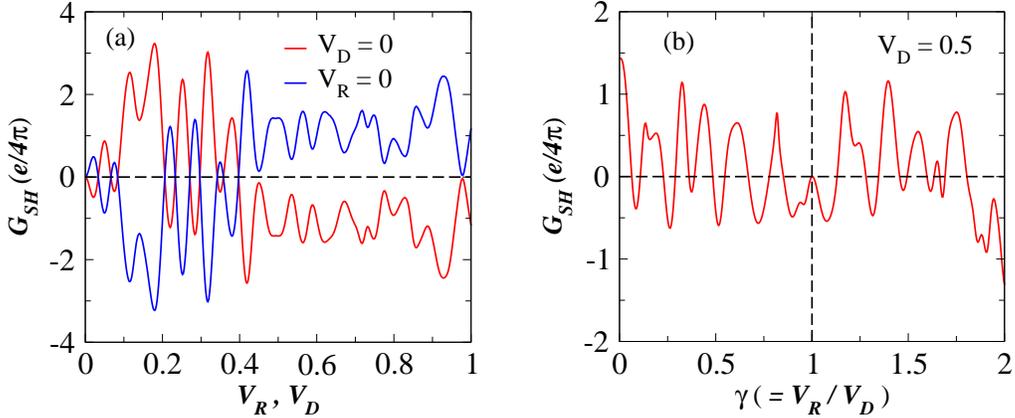

\begin{center}
\includegraphics[width=0.4\textwidth]{fig5a.eps}\quad\quad\includegraphics[width=0.4\textwidth]{fig5b.eps}
\caption{(Color online) (a) Spin Hall conductance, $G_{SH}$ is plotted as a function of $V_R$ (red curve) and $V_D$ (blue curve).   (b) $G_{SH}$ is plotted as a function of the ratio, $\gamma$ $(=V_R/V_D)$. For $\gamma=0$, $G_{SH}$ is zero. Also the angular separation $\theta_Y$ is taken to be $90^\circ$ here.}
\label{vr-vd}
\end{center}
\end{figure}

Let us now study the behaviour of spin Hall conductance as a function of Rashba and Dresselhaus spin-orbit interaction strengths, $V_R$ and $V_D$ respectively. In this case, we set the spin quantization axis along $z$ direction, that is, we fixed $(\theta$, $\phi)$ to $\theta=90^\circ$ and $\phi=0$. Fig.\ref{gs_vr-vd} shows the variation of $G_{SH}$ as a function of $V_R$ and $V_D$. Along the $V_R=V_D$ line, $G_{SH}$ is zero, as seen by the black line and the behaviour of $G_{SH}$ is antisymmetric with respect to the $V_R=V_D$ line. It should be noted that this behaviour is expected, because for $V_R=V_D$, a unitary transformation of the type, $\sigma_x\rightarrow\sigma_y$, $\sigma_y\rightarrow\sigma_x$ and $\sigma_z\rightarrow-\sigma_z$, the Rashba and the Dresselhaus terms get interchanged. By symmetry arguments, the SHC should be zero \cite{shen} (also see Fig.\ref{vr-vd}(a)).

For the three different angles, $\theta_Y$, there is an interesting feature if we look at the order of magnitude of $G_{SH}$. For $\theta_Y=53.13^\circ$, $G_{SH}$ has the maximum value, while $G_{SH}$ is minimum for $\theta_Y=128.87^\circ$. This is because different values of $\theta_Y$, causes different scattering environment for the electrons flowing through the leads 2 and 3. In this regard, the positions of the nearest neighbouring sites also play an important role. Since we are measuring the spin current at terminal 2, for a lower $\theta_Y$, electrons reach terminal 2 more easily in comparison to larger values of $\theta_Y$. In other words, the probability of getting scattered towards terminal 2 will be less for larger values of $\theta_Y$. This explains $G_{SH}$ to be small for $\theta_Y=128.87^\circ$ compared to $\theta_Y=53.13^\circ$ and $\theta_Y=90^\circ$. 

Motivated by the experiments done on semiconductor quantum wells \cite{ganichev}, where the realistic values of the ratio, $\gamma$ $(=V_R/V_D)$ was discussed to be in the range $\sim1.5-2.5$, we have studied the variation of $G_{SH}$ as a function of $V_R$ with $V_D=0$ and vice versa to understand the effects of solely one type of SO interaction.The different parameters are taken as, $E=-2t$, $\theta_Y=90^\circ$ and the spin quantization is aligned along the $z$-axis $(\theta=90^\circ, \phi=0)$ as shown in Fig.\ref{vr-vd}(a) . There is a nice symmetry (differing by a negative sign) among the behaviour of SHC for the Rashba and Dresselhaus interactions. Also in Fig.\ref{vr-vd}(b), we plot $G_{SH}$ as a function of $\gamma$ for a fixed $V_D$, that is, $V_D=0.5$. For $V_R=V_D$, that is, $\gamma=1$, $G_{SH}=0$ as explained earlier. $G_{SH}$ is seen to oscillate about its zero value. The qualitative behaviour of $G_{SH}$ as presented in Fig.\ref{vr-vd} remains unchanged for a different $\theta_Y$ or for other values of the Fermi energy.
\section{Summary and Conclusions}
In summary, in the present work we have studied the effect of the
angular separation of a three terminal Y-shaped junction device in
presence of Rashba and Dresselhaus spin orbit couplings on the spin
Hall conductance by Landauer-B\"{u}ttiker formalism. A prescription for the fabrication of the Y-shaped structures with different angular separation is presented. In presence of RSOC, the colour maps of the spin Hall conductance show interesting features as a function of the parameters denoting the spin quantization axes $(\theta,\phi)$ for three different angular separation of the Y-shaped device. A lower angular separation yields a larger $G_{SH}$ owing to enhanced spin Hall current. The results reveal that the rotational symmetry in spin space is lost owing to the SO couplings present therein. A comparison between RSOC and the SO interaction of the other kind, that is, the Dresselhaus SO interaction is made via studying the behaviour of $G_{SH}$. $G_{SH}$ is antisymmetric in nature with respect to the $V_R=V_D$ line and for $V_R=V_D$, $G_{SH}$ is exactly zero, results that are expected. 

We believe that with the advent of improved fabrication technologies, our studies of SHE in three terminal Y-shaped junction devices can be experimentally achievable and should be instrumental in designing newer spintronic devices.

%\acknowledgments
%Insert here the text.

\end{document}